\documentclass[11pt,twoside]{article}

\usepackage{asp2006}
\usepackage{epsf}
\usepackage{psfig}
\usepackage{lscape}

\usepackage{txfonts,natbib,rotating,amssymb,graphicx} 

\def\amin{^\prime}
\def\adeg{^{\circ}}
\def\Porb{P_{orb}}
\def\microns{\mbox{ } \mu \mbox{m}}
\def\Rstar{\mbox{ }R_{\star}}
\def\Rsol{\mbox{ }R_{\odot}}
\def\Msol{\mbox{ }M_{\odot}}
\def\Lsol{\mbox{ }L_{\odot}}
\def\Mdot{\frac{dM}{dt}}
\def\Tstar{\mbox{ }T_{\star}}
\def\ergs{\mbox{ erg\,s}^{-1}}
\def\nh{N_{\rm H}}
\def\cmmoinsdeux{\mbox{ cm}^{-2}}
\def\kms{\mbox{ km\,s}^{-1}}
\def\mags{\mbox{ magnitudes}}

\begin{document}
\title{High Energy Phenomena in Supergiant X-ray Binaries} 
\author{S. Chaty}   
\affil{Laboratoire AIM, CEA/DSM - CNRS - Universit\'e Paris Diderot,
Irfu/Service d'Astrophysique, Centre de Saclay, B\^at. 709,
F-91191 Gif-sur-Yvette Cedex, France}    

\begin{abstract} 
The INTEGRAL satellite has revealed a major population of supergiant
High Mass X-ray Binaries in our Galaxy, revolutionizing our
understanding of binary systems and their evolution. This population,
constituted of a compact object orbiting around a massive and luminous supergiant
star, exhibits unusual properties, either being extremely absorbed, or
showing very short and intense flares.  An intensive set of
multi-wavelength observations has led us to reveal their nature, and
to show that these systems are wind-fed accretors, closely related to
massive star-forming regions.  In this paper I describe the
characteristics of these sources, showing that this newly
revealed population is linked to the evolution of gamma-ray
emitting massive stars with a compact companion.
\end{abstract}

\section{The $\gamma$-ray sky seen by the {\it INTEGRAL} satellite}

The {\it INTEGRAL} observatory is an ESA satellite launched on 17 October
2002 by a PROTON rocket on an excentric orbit. It is hosting 4
instruments: 2 $\gamma$-ray coded-mask telescopes --the imager IBIS
and the spectro-imager SPI, observing in the range 10 keV-10 MeV, with
a resolution of $12\amin$ and a field-of-view of $19\adeg$-- a
coded-mask telescope JEM-X (3-100 keV), and an optical telescope
(OMC).


The $\gamma$-ray sky seen by {\it INTEGRAL} is very rich, since 499 sources
have been detected by {\it INTEGRAL}, reported in the $3^{rd}$ IBIS/ISGRI soft
$\gamma$-ray catalogue, spanning 3.5 years of observations in the 20-100
keV domain \citep{bird:2007}.  214 sources were discovered by
{\it INTEGRAL}, while the remaining 285 were already known.
Among these sources, there are 147 XRBs (representing 29\% of the
whole sample of sources detected by {\it INTEGRAL}, called ``IGRs'' in the following), 163 AGNs (33\%),
27 CVs (5\%), and 20 sources of other type (4\%): 12 SNRs, 2 globular
clusters, 2 SGRs and 1 GRB. 129 objects still remain unidentified
(26\%).  XRBs are separated in 82 LMXBs and 78 HMXBs (each category
represents 16\% of IGRs).
Among the HMXBs, there are 24 BeXBs and 19 sgXBs (representing
respectively 31\% and 24\% of HMXBs).

It is interesting to follow the evolution of the ratio between BeXBs
and sgXBs.  During the pre-{\it INTEGRAL} era, HMXBs were mostly BeXBs
systems.  For instance, in the catalogue of 130 HMXBs by
\cite{liu:2000}, there were 54 BeXBs and 7 sgXBs (respectively 42\% and 5\% of the total number of
HMXBs).  Then, the situation changed with the first
HMXBs identified by {\it INTEGRAL}: in the catalogue of 114 HMXBs (+128 in Magellanic Clouds) of \cite{liu:2006}, there
were 60\% of BeXBs and 32\% of sgXBs firmly identified.  Therefore,
while the ratio of BeXBs/HMXBs increased by a factor of 1.5 only, the
one of sgXBs/HMXBs increased by a factor of 6.


\section{Let the  {\it INTEGRAL} show go on!}

The ISGRI detector on the IBIS imager has performed a detailed survey of the
Galactic plane, discovering
many new high energy celestial objects, most of which 
reported in \cite{bird:2007}\footnote{See an updated list at 
{\em http://irfu.cea.fr/Sap/IGR-Sources/}}.  The
most important result of {\it INTEGRAL} to date is the discovery of
many new high energy sources -- concentrated in the Galactic plane,
mainly towards tangential directions of Galactic arms, rich in star forming
regions, -- exhibiting common characteristics which previously had
rarely been seen (see e.g. \citeauthor{chaty:2005a}
\citeyear{chaty:2005a}). 
Many of them are HMXBs hosting a NS orbiting around an OB companion, in
most cases a supergiant star. 
Nearly all the {\it INTEGRAL} HMXBs for which both spin and orbital
periods have been measured are located in the upper part of the Corbet
diagramme \citep{corbet:1986}.  They are wind accretors, typical of
supergiant HMXBs, and X-ray pulsars exhibiting longer pulsation
periods and higher absorption (by a factor $\sim4$) as compared to the
average of previously known HMXBs \citep{bodaghee:2007}. 
They divide into two classes: some are very obscured, exhibiting a huge intrinsic and
local extinction, --the most extreme example being the highly absorbed
source IGR~J16318-4848 \citep{filliatre:2004}--, and the others are
HMXBs hosting a supergiant star and exhibiting fast and transient
outbursts -- an unusual characteristic among HMXBs.  These are
therefore called Supergiant Fast X-ray Transients (SFXTs,
\citeauthor{negueruela:2006a} \citeyear{negueruela:2006a}),
with IGR~J17544-2619
being their archetype \citep{pellizza:2006}.

\subsection{Multi-wavelength observations of  {\it INTEGRAL} sources} \label{observations-IGRs}

To better characterise this population, \cite{chaty:2008} and \cite{rahoui:2008} 
studied a sample of 21 IGRs
belonging to both classes described above.  Sources of
this sample are X-ray pulsars, with high $P_\mathrm{spin}$ from 139 to 5880\,s
and $\Porb$ ranging from 4 to 14\,days.  They are therefore wind
accreting supergiant HMXBs, according to the Corbet diagramme.  The
multiwavelength observations were performed from 2004 to 2008 at the
European Southern Observatory (ESO), using Target of Opportunity (ToO)
and Visitor modes, in 3 domains: optical ($400-800$\,nm) with EMMI,
NIR ($1-2.5 \microns$) with SOFI, both instruments at the focus of the
3.5m New Technology Telescope (NTT) at La Silla, and mid-infrared
(MIR, $5-20 \microns$) with the VISIR instrument on Melipal, the 8m
Unit Telescope 3 (UT3) of the Very Large Telescope (VLT) at Paranal
(Chile). They also used data from the GLIMPSE survey of {\it Spitzer}.  With
these observations they performed accurate astrometry, identification,
photometry and spectroscopy on this sample of IGRs,
aiming at identifying their counterparts and the nature of the
companion star, deriving their distance, and finally characterising
the presence and temperature of their circumstellar medium, by fitting
their spectral energy distribution (SED).  

The main results of this study are that 15 of
these IGRs are identified as HMXBs, and among them 12 HMXBs contain
massive and luminous early-type companion stars. By combining optical,
NIR and MIR photometry, and fitting their SEDs, \cite{rahoui:2008}
showed that (i) most of these sources exhibit an intrinsic absorption
and (ii) three of them exhibit a MIR excess, which they suggest to be
due to the presence of a cocoon of dust and/or cold gas enshrouding
the whole binary system, with a temperature of $T_d \sim 1000$\,K,
extending on a radius of $R_d \sim 10\Rstar$ (see \citeauthor{chaty:2006c} \citeyear{chaty:2006c}).

\subsection{Supergiant Fast X-ray Transients}

\subsubsection{General characteristics}

SFXTs constitute a new class of $\sim 12$ sources identified among the
recently discovered IGRs. They are HMXBs hosting NS orbiting around sgOB companion stars, exhibiting peculiar characteristics compared to ``classical'' HMXBs: rapid outbursts lasting only for hours, faint quiescent emission, and high energy spectra requiring a BH or NS accretor. The flares rise in tens of minutes, last for $\sim$ 1 hour, their frequency is $\sim7$\,days, and their luminosity $L_x \sim 10^{36} \ergs$ at the outburst peak.

   \subsubsection{IGR~J17544-2619, archetype of SFXTs}

\begin{figure}
  \includegraphics[height=.35\textheight,angle=-90]{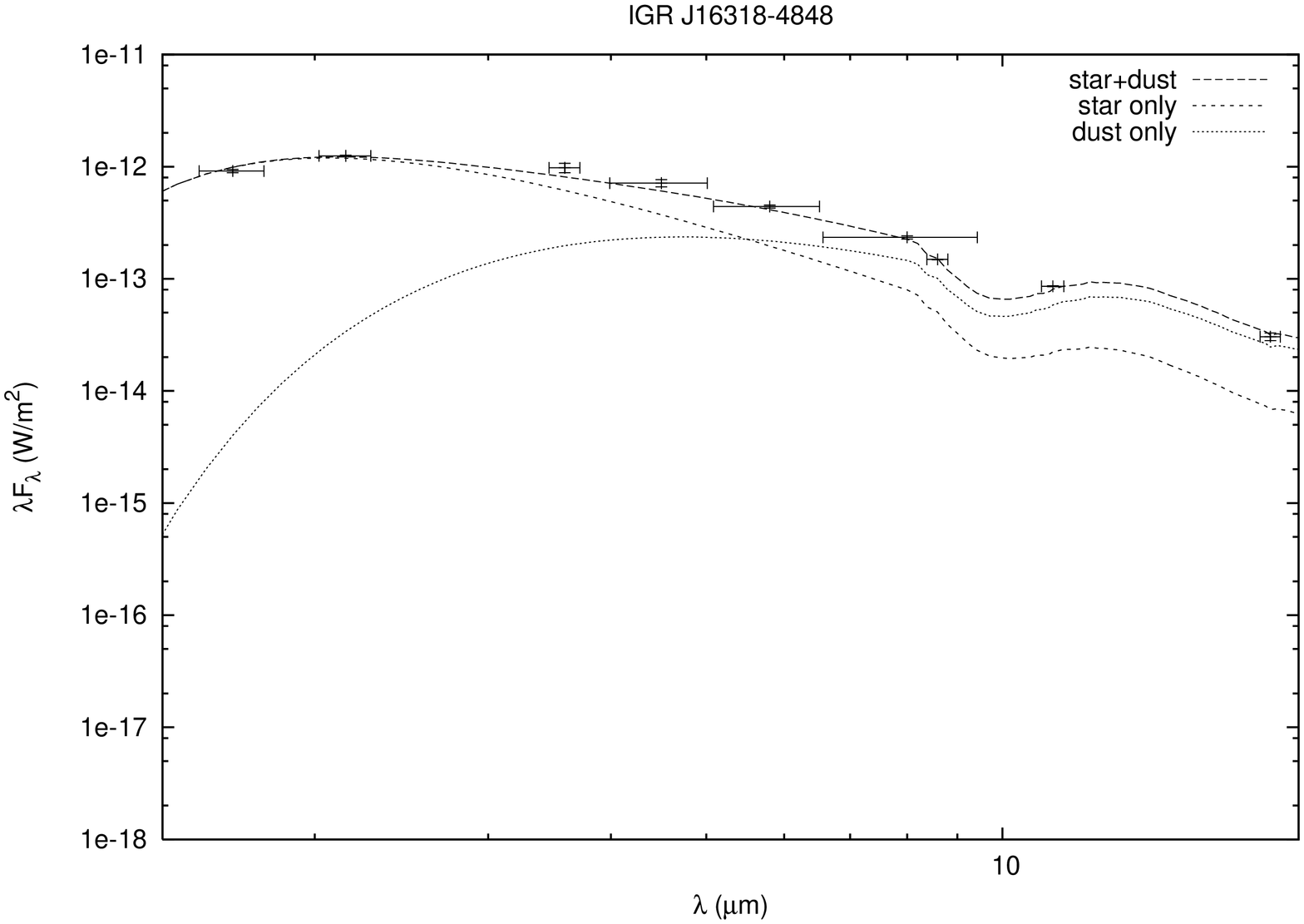}
  \includegraphics[height=.37\textheight,angle=-90]{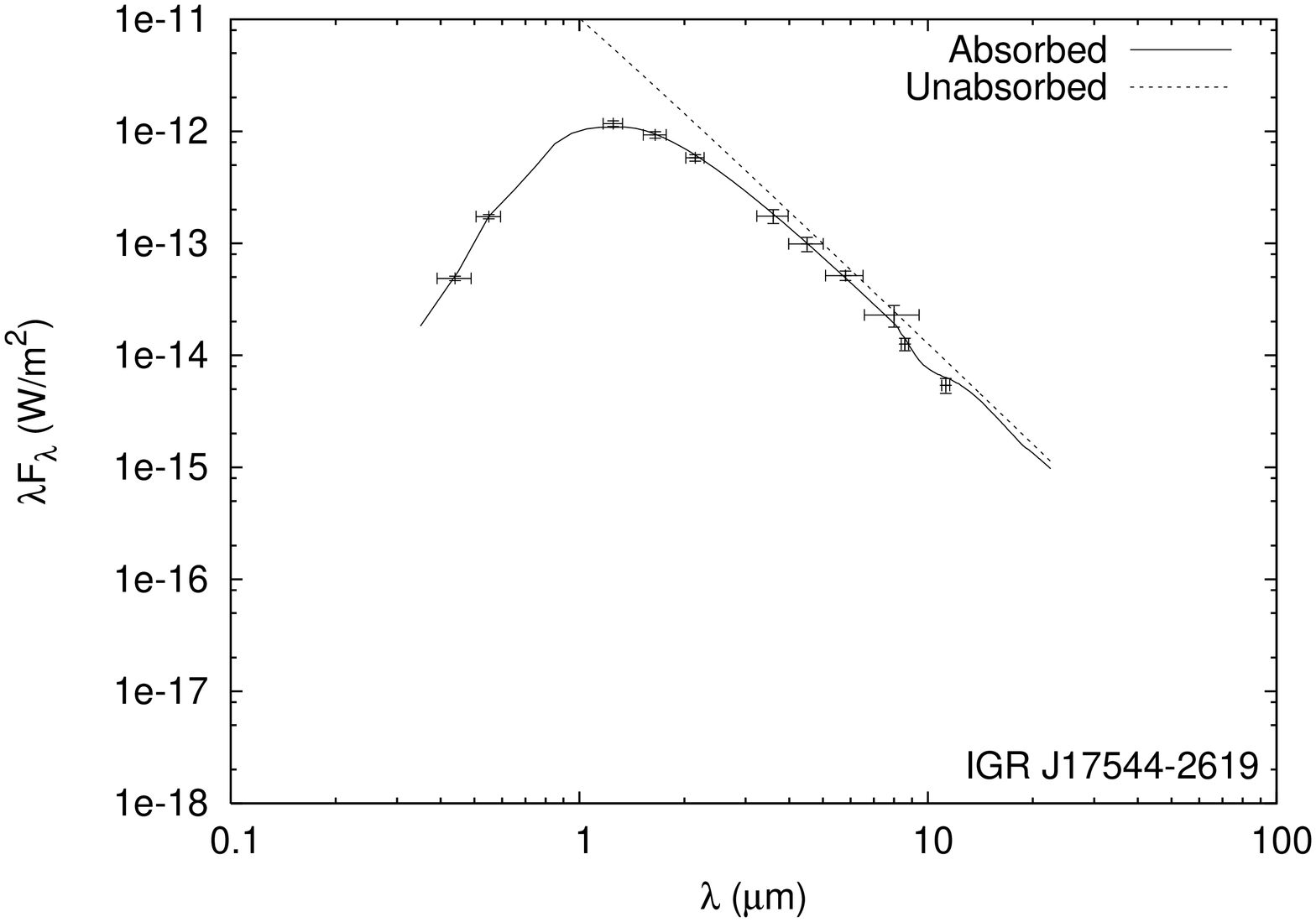}
  \caption{\label{figure:igrj16318-igrj17544} Optical to MIR SEDs of
  IGR~J16318-4848 (left) and IGR~J17544-2619 (right), including data
  from ESO/NTT, VISIR on VLT/UT3 and {\it Spitzer} \citep{rahoui:2008}.
  IGR~J16318-4848 exhibits a MIR excess, interpreted as the signature of a strong stellar outflow
  coming from the sgB[e] companion star \citep{filliatre:2004}.  On the
  other hand, IGR~J17544-2619 is well fitted with only a stellar
  component corresponding to the O9Ib companion star spectral type
  \citep{pellizza:2006}.}
\end{figure}

   This bright recurrent transient X-ray source was discovered by {\it
     INTEGRAL} on 17 September 2003 \citep{sunyaev:2003b}. {\it
     XMM-Newton} observations showed that it exhibits a very hard
   X-ray spectrum, and a relatively low intrinsic absorption ($\nh
   \sim 2 \times 10^{22}\cmmoinsdeux$,
   \citeauthor{gonzalez-riestra:2004}
   \citeyear{gonzalez-riestra:2004}).  Its bursts last for hours, and
   inbetween bursts it exhibits long quiescent periods, which can
   reach more than 70\,days. The X-ray behaviour is complex on long,
   mean and short-term timescales: rapid flares are detected by {\it
     INTEGRAL} on all these timescales, on pointed and 200s binned
   lightcurve (Zurita Heras \& Chaty in prep.). The compact object is
   probably a NS \citep{intzand:2005}.  \cite{pellizza:2006} managed
   to get optical/NIR ToO observations only one day after the
   discovery of this source. They identified a likely counterpart
   inside the {\it XMM-Newton} error circle, confirmed by an accurate
   localization from {\it Chandra}.  Spectroscopy showed that the
   companion star was a blue supergiant of spectral type O9Ib, with a
   mass of $25-28 \Msol$, a temperature of $T\sim 31000$~K, and a
   stellar wind velocity of $265 \pm 20 \kms$ (which is faint for O
   stars): the system is therefore an HMXB \citep{pellizza:2006}.
   \cite{rahoui:2008} combined optical, NIR and MIR observations and
   showed that they could accurately fit the observations with a model
   of an O9Ib star, with a temperature $\Tstar \sim 31000$~K and a
   radius $\Rstar = 21.9 \Rsol$. They derived an absorption A$_v = 6.1
   \mags$ and a distance D~$=3.6$~kpc. Therefore the source does not
   exhibit any MIR excess, it is well fitted by a unique stellar
   component (see Figure \ref{figure:igrj16318-igrj17544}, right
   panel, \citeauthor{rahoui:2008} \citeyear{rahoui:2008}).

\subsubsection{Classification of SFXTs}

We can divide the SFXTs in two groups, according to the duration and
frequency of their outbursts, and their
$\frac{L_\mathrm{max}}{L_\mathrm{min}}$ ratio.  The classical SFXTs
exhibit a very low quiescence $L_X$ and a high variability, while
intermediate SFXTs exhibit a higher $<L_X>$, a lower
$\frac{L_\mathrm{max}}{L_\mathrm{min}}$ and a smaller variability,
with longer flares.  SFXTs might appear like persistent sgXBs with
$<L_X>$ below the canonical value of $\sim 10^{36} \ergs$, and flares
superimposed.  But there might be some observational bias in these
general characteristics, therefore the distinction between SFXTs and
sgXBs is not well defined yet.
While the typical hard X-ray variability factor (the ratio between
deep quiescence and outburst flux) is less than 20 in
classical/absorbed systems, it is higher than 100 in SFXTs (some
sources can exhibit flares in a few minutes, like for instance
XTE\,J1739-302 \& IGR\,J17544-2619).  The
intermediate SFXTs exhibit smaller variability factors. \\

\underline{SFXT behaviour: clumpy wind accretion?}

Such sharp rises exhibited by SFXTs are incompatible with the orbital
motion of a compact object through a smooth medium
(\citeauthor{negueruela:2006a} \citeyear{negueruela:2006a},
\citeauthor{smith:2006} \citeyear{smith:2006},
\citeauthor{sguera:2005} \citeyear{sguera:2005}).  Instead, flares
must be created by the interaction of the accreting compact object
with the dense clumpy stellar wind (representing a large fraction of
stellar $\Mdot$).  In this
case, the flare frequency depends on the system geometry, and the
quiescent emission is due to accretion onto the compact object of
diluted inter-clump medium, explaining the very low quiescence level
in classical SFXTs. \\

\underline{Macro-clumping scenario}

Each SFXT outburst is due to the accretion of a single clump, assuming that 
the X-ray lightcurve is a direct tracer of the wind density distribution.
The typical parameters in this scenario are:
a compact object with large orbital radius: $10 \Rstar$,
a clump size of a few tenths of $\Rstar$,
a clump mass of $10^{22-23}g$ (for $\nh=10^{22-23}\cmmoinsdeux$),
a mass loss rate of $10^{-(5-6)} \Msol/yr$,
a clump separation of order $R_{\star}$ at the orbital radius,
and a volume filling factor: 0.02$->$0.1.
The flare to quiescent count rate ratio is directly related to the $\frac{clump}{inter-clump}$ density ratio, which ranges between
15-50 for intermediate SFXTs, and $10^{2-4}$ for "classical" SFXTs.
A very high degree of porosity (macroclumping) is required to reproduce
the observed outburst frequency in SFXTs, in good agreement with UV line
profiles and line-driven instabilities at large radii 
(\citeauthor{oskinova:2007} \citeyear{oskinova:2007};
\citeauthor{runacres:2005} \citeyear{runacres:2005}; 
\citeauthor{walter:2007} \citeyear{walter:2007}). \\

\underline{Difference sgXB/SFXT}

To explain the emission of sgXB/SFXT, 
\cite{negueruela:2008} and \cite{walter:2007} invoke the
existence of two zones around the supergiant star, of high and low
clump density respectively.  This would naturally explain the smooth
transition between sgXBs and SFXTs, and the existence of intermediate
systems; the main difference between classical sgXBs and SFXTs being
in this scenario the NS orbital radius.
Indeed, a basic model of porous wind predicts a substantial change in the
properties of the wind "seen by the NS" at a distance $r \sim 2
\Rstar$ (\citeauthor{negueruela:2008} \citeyear{negueruela:2008}), where we
stop seeing persistent X-ray sources. There are 2-regimes:
either the NS sees a large number of clumps, because it is
  embedded in a quasi-continuous wind;
or the number density of clumps is so small that the NS is effectively 
orbiting in an empty space.

The observed division between sgXBs (persistent sgXBs and SFXTs)
is therefore naturally explained by simple geometrical differences
in the orbital configurations:

1. The obscured sgXBs (persistent and luminous systems) would have 
short and circular orbits lying 
inside the zone of stellar wind high clump density ($R_{orb} \sim 2\Rstar$).

2. The intermediate SFXTs would have short orbits, circular or eccentric, 
and possible periodic outbursts, the NS being inside the narrow transition zone.

3. The classical SFXTs would have larger and eccentric orbital radius,
the NS orbiting outside the high density zone. \\

\underline{IGR\,J18483-0311: an intermediate SFXT?}

X-ray properties of this system were suggesting an SFXT
\citep{sguera:2007}, exhibiting however an unusual behaviour: its
outbursts last for a few days (to compare to hours for classical
SFXTs), and the ratio $L_{max}/L_{min} \sim 10^3$ (its quiescence is
therefore at a higher level than the ratio $\sim 10^4$ for classical
SFXTs). Moreover, its orbital period $\Porb$=18.5d is low compared to
classical SFXTs (with large/eccentric orbits). Finally, its orbital
and spin periods ($P_\mathrm{spin}$=21.05s) located it ambiguously inbetween Be and
sgXBs in the Corbet Diagramme.
\cite{rahoui:2008a} identified the companion star of this system
as a B0.5Ia supergiant, unambiguously showing that this system is an
SFXT.  Furthermore, they suggest that this system could be the first
firmly identified intermediate SFXT, characterised by short, eccentric
orbit (with an eccentricity $e$ between 0.4 and 0.6),
and long outbursts... An "intermediate"
SFXT nature would explain the unusual characteristics of this source among
"classical" SFXTs.

\subsection{Obscured HMXBs}

  \subsubsection{IGR~J16318-4848, an extreme case}

  IGR~J16318-4848 was the first source discovered by IBIS/ISGRI on
  {\it INTEGRAL} on 29 January 2003 \citep{courvoisier:2003}, with a
  $2 \amin$ uncertainty.  {\it XMM-Newton} observations revealed a
  comptonised spectrum exhibiting an unusually high level of
  absorption: $\nh \sim 1.84 \times 10^{24} \cmmoinsdeux$
  \citep{matt:2003}.  The accurate localisation by {\it XMM-Newton}
  allowed \cite{filliatre:2004} to rapidly trigger ToO photometric and
  spectroscopic observations in optical/NIR, leading to the
  confirmation of the optical counterpart \citep{walter:2003} and to
  the discovery of the NIR one \citep{filliatre:2004}.  The extremely
  bright NIR source (B\,$>25.4\pm1$; I\,$=16.05\pm0.54$, J\,$= 10.33\pm 0.14$;
  H\,$=8.33\pm 0.10$ and Ks\,$=7.20 \pm 0.05 \mags$) exhibits an
  unusually strong intrinsic absorption in the optical ($A_v = 17.4
  \mags$), 100 times stronger than the interstellar absorption along
  the line of sight ($A_v = 11.4 \mags$), but still 100 times lower
  than the absorption in X-rays.  This led \cite{filliatre:2004} to
  suggest that the material absorbing in X-rays was concentrated
  around the compact object, while the material absorbing in
  optical/NIR was enshrouding the whole system.  The NIR spectroscopy
  in the $0.95-2.5 \microns$ domain allowed them to identify the nature of
  the companion star, by revealing an unusual spectrum, with many
  strong emission lines:

$\bullet$ H, He${\rm I}$ (P-Cyg) lines, characteristic of dense/ionised wind at v\,$=400$\,km/s,

$\bullet$ He${\rm II}$ lines: the signature of a highly excited region,

$\bullet$ $[$Fe${\rm II}]$: reminiscent of shock heated matter,

$\bullet$ Fe${\rm II}$: emanating from media of densities $>10^5-10^6$\,cm$^{-3}$,

$\bullet$ Na${\rm I}$: coming from cold/dense regions.

All these lines originate from a highly complex, stratified
circumstellar environment of various densities and temperatures,
suggesting the presence of an envelope and strong stellar outflow
responsible for the absorption. Only luminous early-type stars such as
sgB[e] show such extreme environments, and
\cite{filliatre:2004} concluded that IGR~J16318-4848 was an unusual
HMXB hosting a sgB[e] with characteristic luminosity of
$10^6 \Lsol$ and mass of $30 \Msol$, located at a distance
between 1 and 6 kpc (see also \citeauthor{chaty:2005a} \citeyear{chaty:2005a}).
This source would therefore be the second HMXB hosting a sgB[e] star,
after CI Cam (see \citeauthor{clark:1999} \citeyear{clark:1999}). 

The question of this huge absorption was still pending, 
and only MIR observations would allow to solve this question,
and understand its origin.
By combining optical, NIR and MIR observations, and
fitting these observations with a model of sgB[e] companion star,
\cite{rahoui:2008} showed that IGR~J16318-4848 was exhibiting a MIR
excess (see Figure \ref{figure:igrj16318-igrj17544}, left panel), that they
interpreted as due to the strong stellar outflow emanating from
the sgB[e] companion star.  They found that the companion star had a
temperature of $\Tstar=22200$\,K and radius $\Rstar = 20.4 \Rsol = 0.1$\,a.u., 
consistent with a supergiant star, and
an extra component of temperature T $=1100$\,K and radius R\,$= 11.9\Rstar 
= 1$\,a.u., with A$_v = 17.6 \mags$. 
%
%
Recent MIR spectroscopic observations with VISIR at the VLT showed
that the source was exhibiting strong emission lines of H, He, Ne, PAH, Si,
proving that the extra absorbing component was made of dust and
cold gas.

By taking a typical orbital period of 10\,days and a mass of the
companion star of $20 \Msol$, we obtain an orbital separation of $50
\Rsol$, smaller than the extension of the extra component of dust/gas
($= 240 \Rsol$),
suggesting that this dense and absorbing circumstellar material envelope enshrouds the whole binary system, like 
a cocoon (see Figure \ref{figure:obscured-sfxt}, left panel).
We point out that this source exhibits such extreme
characteristics that it might not be fully representative of the other
obscured sources.

\subsection{The Grand Unification: different geometries, different scenarios}

In view of the results described above, 
there seems to be a continuous trend, from classical and/or absorbed
sgHMBs, to classical SFXTs. We outline in the following this trend.

\begin{figure*}[!ht]
\centering
\includegraphics[height=.31\textheight,angle=-90]{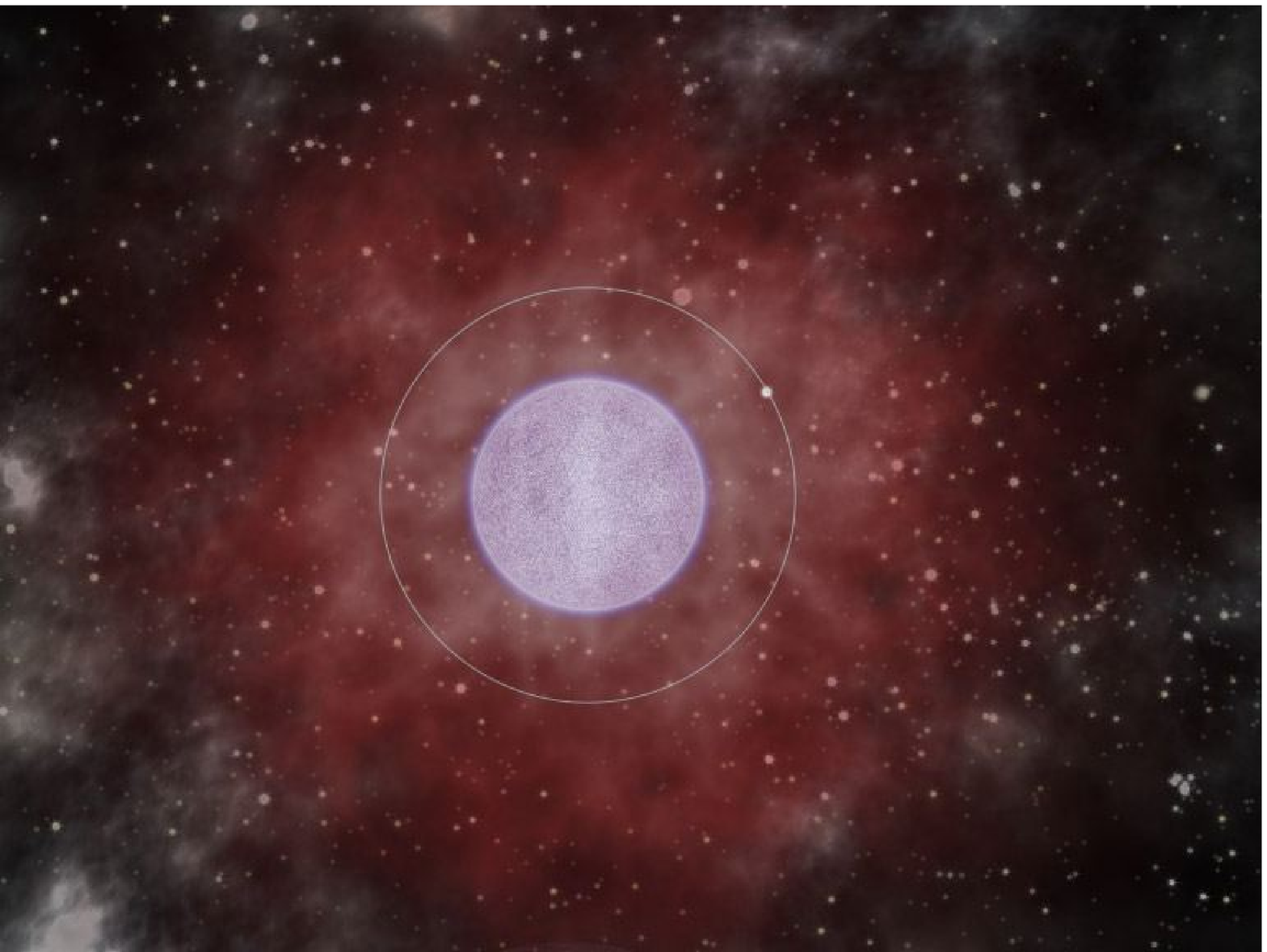}
\includegraphics[height=.31\textheight,angle=-90]{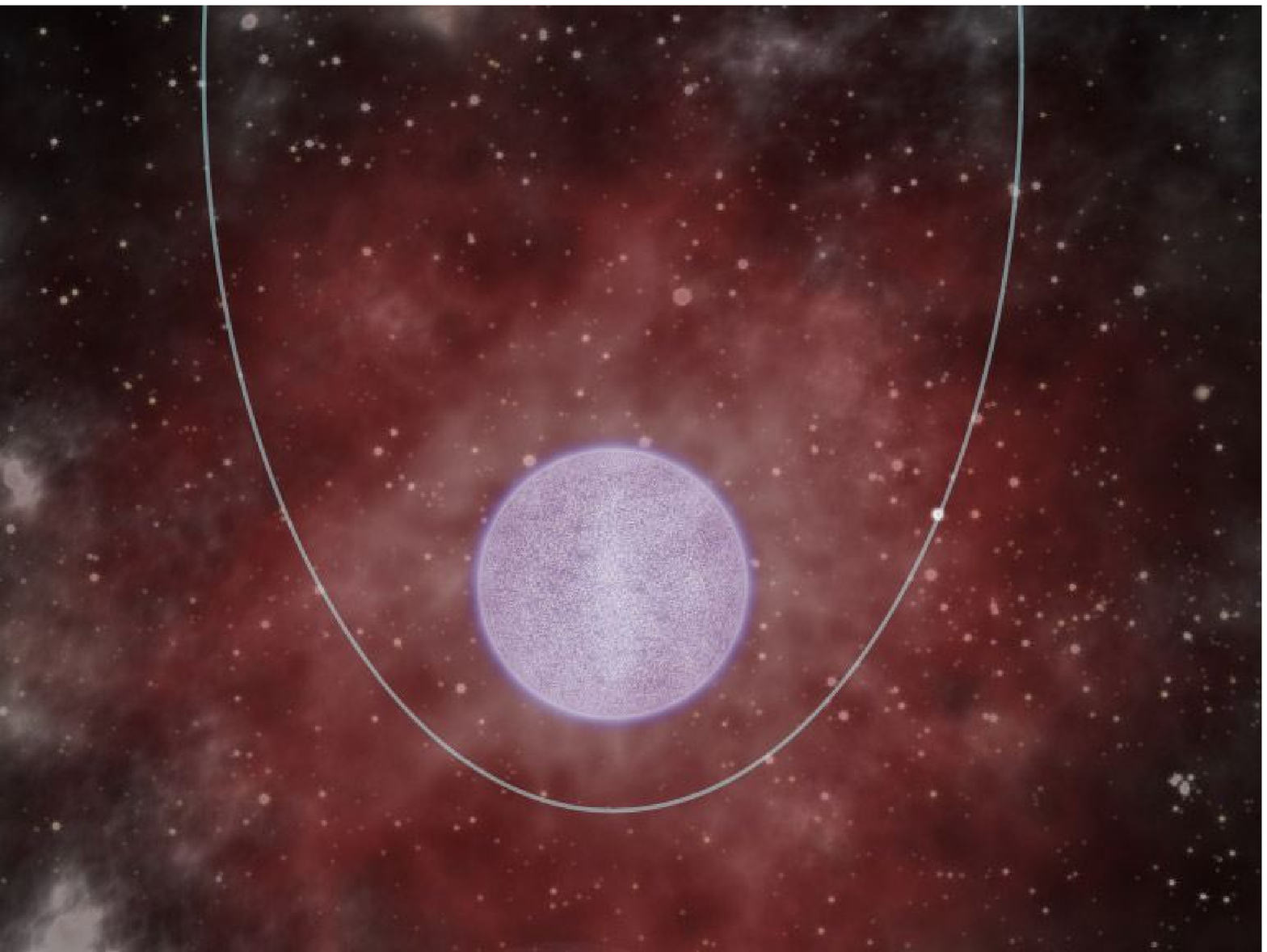}
\caption[Scenarios illustrating both 2 types of {\it INTEGRAL}
sources]{Scenarios illustrating two possible configurations of {\it
    INTEGRAL} sources: a NS orbiting a supergiant
  star on a circular orbit (left image); and on an eccentric orbit
  (right image), accreting from the clumpy stellar wind of the
  supergiant.  The accretion of matter is persistent in the case of
  the obscured sources, as in the left image, where the compact object
  orbits inside the cocoon of dust enshrouding the whole system. On
  the other hand, the accretion is intermittent in the case of SFXTs,
  which might correspond to a compact object on an eccentric orbit, as
  in the right image.  A 3D animation of these sources is available on
  the
  website:\\  
{\em http://www.aim.univ-paris7.fr/CHATY/Research/hidden.html}
}
  \label{figure:obscured-sfxt}
\end{figure*}

1. In "classical" sgXBs, the NS is orbiting at a few
  stellar radii only from the star. The absorbed (or obscured) sgXBs (like
  IGR\,J16318-4848) are classical sgXBs hosting NS constantly
  orbiting inside a cocoon made of dust and/or cold gas, probably
  created by the companion star itself. These systems therefore exhibit a
  persistent X-ray emission.  The cocoon, with an extension of $\sim
  10 \Rstar = 1$\,a.u., is enshrouding the whole binary system. The NS
  has a small and circular orbit (see Figure
  \ref{figure:obscured-sfxt}, left panel).

2. In "Intermediate" SFXT systems (such as IGR\,J18483-0311), the
  NS orbits on a small and circular/excentric orbit, and it is only
  when the NS is close enough to the supergiant star that accretion
  takes place, and that X-ray emission arises.

3. In "classical" SFXTs (such as IGR\,J17544-2619), the NS orbits on
  a large and excentric orbit around the supergiant star, and exhibits some
  recurrent and short transient X-ray flares, while it comes close to
  the star, and accretes from clumps of matter coming from the wind of
  the supergiant.  Because it is passing through more diluted medium,
  the $\frac{Lmax}{Lmin}$ ratio is higher for "classical" SFXTs than for
  "intermediate" SFXTs (see Figure \ref{figure:obscured-sfxt}, right
  panel).


Although this scenario seems to describe quite well the characteristics
currently seen in sgXBs, we still need to identify the nature of
many more sgXBs to confirm it, and in particular the
orbital period and the dependance of the column density with the phase
of the binary system.

\subsubsection{Population synthesis models}

These sources revealed by {\it INTEGRAL}, namely the supergiant HMXBs, will
allow to better constrain and understand the
formation and evolution of binary systems, by comparing them to
numerical study of LMXB/HMXB population synthesis models. 
For instance, these
new systems might represent a precursor stage of what is known as the
"Common envelope phase" in the evolution of LMXB/HMXB systems.
 Many
parameters do influence the various evolutions of these systems:
differences in size, orbital period, ages, accretion type, and stellar
endpoints... Moreover, stellar and circumstellar properties also
influence the evolution of high-energy binary systems, made of two
massive components usually born in rich star forming regions.  
We still have to identify black holes orbiting around supergiant companion stars in wind-accreting HMXBs, however this is only feasible through observational methods involving detection of extremely faint radial velocity displacement due to the high mass of the companion star.
Finally, these sources are also useful to look for massive stellar
"progenitors", for instance giving birth to coalescence of compact
objects, through NS/NS or NS/BH collisions. They would then become
prime candidate for gravitational wave emitters, or even to short/hard
$\gamma$-ray bursts.

\section{Conclusions and perspectives...}

The {\it INTEGRAL} satellite has tripled the total number of Galactic
sgXBs, constituted of a NS orbiting around a supergiant
star. Most of these new sources are slow and absorbed X-ray pulsars,
exhibiting a large $\nh$ and long $P_\mathrm{spin}$ ($\sim$1ks).  The
influence of the local absorbing matter on periodic modulations is
different for sgOB or BeXBs, segregated in
different parts of $\nh$-$\Porb$ or $\nh$-$P_\mathrm{spin}$.
%
{\it INTEGRAL} revealed 2 new types of sources.  First, the SFXTs, exhibiting short and strong X-ray
flares, with a peak flux of 1 Crab during 1--100s, every $\sim 100$\,days.
These flares can be explained by accretion through
clumpy winds.  Second, the obscured HMXBs are persistent X-ray sources
composed of supergiant stellar companions exhibiting a strong intrinsic absorption and long $P_\mathrm{spin}$.  The NS is
deeply embedded in the dense stellar wind, forming a dust cocoon
enshrouding the whole binary system.

These results show the existence in our Galaxy of a dominant
population of a previously rare class of high-energy binary systems:
supergiant HMXBs, some exhibiting a high intrinsic absorption
(\citeauthor{chaty:2008} \citeyear{chaty:2008};
\citeauthor{rahoui:2008} \citeyear{rahoui:2008}). Studying
this population will provide a
better understanding of the formation and evolution of short-living
HMXBs.  Furthermore, stellar population models now have to
take these objects into account, to assess a realistic number of
high-energy binary systems in our Galaxy. 



\acknowledgements 
I thank the organisers for such a successfully organized and nice workshop,
that I would like to dedicate to the new-born star Carolina Mart{\'i}!




\end{document}